\documentclass[aip,prb,twocolumn,showpacs]{revtex4}
\usepackage{graphicx,color}
\usepackage{amsmath}
\usepackage{longtable}
\begin{document}

\title{Ion sphere model for Yukawa systems (dusty plasmas) }

\author{S. A. Khrapak,$^{1,2}$ A. G. Khrapak,$^2$ A. V. Ivlev,$^3$ and H. M. Thomas$^1$
}
\date{\today}
\affiliation{ $^1$Forschungsgruppe Komplexe Plasmen, Deutsches Zentrum f\"{u}r Luft- und Raumfahrt,
Oberpfaffenhofen, Germany\\ $^2$Joint Institute for High Temperatures RAS, Moscow, Russia
\\$^3$Max-Planck-Institut f\"ur extraterrestrische Physik, Garching,Germany
}

\begin{abstract}
Application of the ion sphere model (ISM), well known in the context of the one-component-plasma, to estimate thermodynamic properties of model Yukawa systems is discussed. It is shown that the ISM approximation provides fairly good estimate of the internal energy of the strongly coupled Yukawa systems, in both fluid and solid phases. Simple expressions for the excess pressure and isothermal compressibility are derived, which can be particularly useful in connection to wave phenomena in strongly coupled dusty plasmas. It is also shown that in the regime of strong screening a simple consideration of neighboring particles interactions can be sufficient to obtain quite accurate estimates of thermodynamic properties of Yukawa systems.
\end{abstract}

\pacs{52.27.Lw; 52.25.Kn}
\maketitle

\section{Introduction}

Thermodynamic properties of Yukawa systems (particles interacting via Yukawa or Debye-H\"{u}ckel pair potential) are of considerable interest, in particular in the context of physics of plasmas, dusty (complex) plasmas, and colloidal dispersions. An idealized model dealing with point-like charges immersed in a neutralizing medium, which is responsible for the exponential screening of the interaction potential, has been extensively investigated using various simulation and analytical techniques. Very accurate results for the thermodynamic functions of this model obtained from Monte-Carlo (MC) and molecular dynamics (MD) simulations~\cite{Meijer,Farouki1994,Hamaguchi96,Hamaguchi,Caillol} and integral equation theories~\cite{Kalman2000,Faussurier} are available in the literature.

The idealized model of Yukawa systems disregards some important properties of real substances. Some of these properties, which are particularly relevant to
dusty plasmas (and to some extent also to colloids) are as follows:~\cite{TsytovichUFN,FortovUFN,FortovPR,KM2009,Bonitz,SAKEPL,ChaudhuriSM,FortovBook,IvlevBook} Particles are not point-like, the typical ratio of the particle size to the plasma screening length can vary in a relatively wide range; There is a wide region around the particles where the ion-particle interaction is very strong, which results in non-linear screening; Plasma electrons and ions are continuously deposited on the particle surface, which results in considerable deviations from the equilibrium (Boltzmann) distribution of these plasma species; Particle charge is not fixed, but depends on various system parameters (e.g. on the density of the particles themselves); The average
density of ions and electrons is not fixed, but is related to the particle density and charge via the quasineutrality
condition. All this complicates direct application of existing results to practical situations. More realistic models to represent real dusty plasmas under various conditions are required.

One of the possible strategies towards such models is to construct simple analytical approximations for the ``basic'' case, corresponding to the idealization discussed above. This can then serve as the basis of more realistic models, allowing an easy evaluation of the relative importance of specific dusty plasma properties in each concrete situation. Existing accurate results for an idealized
Yukawa model can be considered as reference data in constructing such simple analytical approximations.

This point of view has been shared in a previous publication,~\cite{DHH} where the Debye-H\"{u}ckel plus hole (DHH) approximation has been applied to Yukawa systems. The DHH approach has been originally proposed to reduce inaccuracy of the conventional Debye-H\"{u}ckel (DH) theory when evaluating thermodynamic properties of non-ideal one-component-plasma (OCP).~\cite{Nordholm} The main idea behind the DHH approximation is that the exponential particle density must be truncated close to a test particle so as not to become negative upon linearization.
When applied to Yukawa systems, DHH demonstrates considerable improvement over the traditional DH theory in the regime covering the transition between weak and moderate coupling (in the limit of weak coupling DHH reduces to DH). It even allows to roughly reproduce the thermodynamics of strongly coupled Yukawa systems up to the fluid-solid phase transition, but the agreement with available results from numerical experiments is not better than qualitative in this regime.~\cite{DHH}

The purpose of the present paper is to discuss another simple model, which is particularly suitable for the regime of strong coupling. This is the so-called ion sphere model (ISM), which is known to  describe rather precisely the internal energy of the OCP in the limit of strong coupling.~\cite{Baus,Ichimaru,Dubin} We show that ISM allows to obtain simple expressions for the internal energy, pressure, and compressibility of Yukawa systems. These expressions are compared with the ``exact'' reference data from MD simulations.~\cite{Farouki1994,Hamaguchi96,Hamaguchi}
Good agreement is found in the regime of strong coupling, especially in the weakly screened regime. Overall, the ISM approximation is shown to be more simple and more accurate than DHH at strong coupling. As such, it provides a good basis for developing more realistic models to describe thermodynamic properties of dusty plasmas and related systems under various natural and laboratory conditions.

Following our previous paper,~\cite{DHH} we adopt the following simplified model.
The two-component system consists of particles of charge $Q$ and density $n_{\rm p}$ immersed into a neutralizing
medium, characterized by the charge $-e$ and density $n$ (relation to the conventional three-component dusty plasma is discussed in Appendix~\ref{ApA}). In equilibrium the system is
quasineutral, so that $Qn_{\rm{p}0} - en_{0}=0$, where the subscript $0$ denotes unperturbed quantities.  The system is characterized by two dimensionless parameters:
\begin{equation}
\Gamma=\frac{Q^2}{a T} ~~~~{\rm and}~~~~ \kappa=ak_{\rm D},
\end{equation}
where $a= (3/4\pi n_{\rm{p}})^{1/3}$ is the Wigner-Seitz radius, $T$ is the temperature (in energy units), and $k_{\rm
D}=\sqrt{4\pi e^2 n_0/T}$ is the inverse screening length (Debye radius) associated with the neutralizing medium.  The coupling parameter $\Gamma$ is roughly the ratio of the Coulomb interaction energy, evaluated at the mean interparticle separation, to the kinetic energy. The screening parameter $\kappa$ is the ratio of the
interparticle separation to the screening length.

The main quantities of interest are the internal energy $U$, Helmholtz free energy $F$, and pressure $P$, associated with the particle component. In reduced units these are
\begin{equation}
u=U/NT,~~~ f=F/NT, ~~~ p=PV/NT,
\end{equation}
where $N$ is the number of particles in the volume $V$ (so that $n_{\rm p}=N/V$).

\section{Ion sphere model and static excess energy}

The ion sphere model for the non-ideal OCP has the following simple physical interpretation.~\cite{Baus,Ichimaru,Dubin} Consider $N$ charged particles immersed in the uniform neutralizing background.
Due to the strong Coulomb repulsive interaction between the particles, they tend to form a regular structure with the interparticle distance of order $a$. One can think of a collection of $N$ particles together with a spherical piece of the uniform background of radius $a$, which exactly compensates the particle charge. It is then assumed that the energy of the system is just the sum of the energies of such spheres. The energy of the sphere can be easily calculated via purely electrostatic arguments, resulting in the celebrated expression
\begin{equation}\label{uOCP}
u_{\rm OCP}= -\tfrac{9}{10}\Gamma.
\end{equation}
The numerical coefficient $-0.9$ is very close to the Madelung constants of the body-centered-cubic (bcc) and face-ceneterd-cubic (fcc) crystals, which are $-0.8959$ and $-0.8958$, respectively.

Some modifications are required when applying these arguments to the Yukawa system. We again divide the system into $N$ charge neutral cells (ion spheres) of radius $a$, with each particle placed in the center of the cell. The electrical potential inside the cell is given by the Poisson equation
\begin{equation}
\Delta \phi = - 4\pi Q\delta({\bf r}) + 4\pi e n.
\end{equation}
The density of the neutralizing medium satisfies the linearized Boltzmann relation
\begin{equation}
n=n_0\left(1+e\phi/T\right).
\end{equation}
The general solution of the Poisson equation has the form
\begin{equation}\label{sol1}
\phi(r)= ({\mathcal A}_1/r)e^{-k_{\rm D}r} + ({\mathcal A}_2/r)e^{k_{\rm D}r} - 3Q/\kappa^2a,
\end{equation}
where ${\mathcal A}_1$ and ${\mathcal A}_1$ are the coefficients to be determined. To do this we use the boundary condition $\phi'(a)=0$, which follows from the cell charge neutrality. The second requirement is ${\mathcal A}_1+{\mathcal A}_2=Q$, implying that $\phi$ tends to $Q/r$ as $r\rightarrow 0$. These conditions yield  ${\mathcal A}_1$ and ${\mathcal A}_2$ and hence the electrical potential inside the cell,
\begin{equation}\label{potential}
\phi(r)=\frac{Q}{r}e^{-k_{\rm D}r}+\frac{Q}{r}\frac{2(\kappa+1)\sinh(k_{\rm D}r)}{(\kappa+1)+(\kappa-1)e^{2\kappa}}-\frac{3Q}{\kappa^2 a},
\end{equation}
where the first term is the conventional Debye-H\"{u}ckel potential of an individual particle in plasma and the last two terms arise due to requirements imposed by the ion sphere model.
The reduced electrostatic energy of the sphere can be calculated from the conventional expression
\begin{equation}\label{energydef}
u_{\rm st}=\frac{1}{2}\frac{Q}{T}\left[\phi(r)-\frac{Q}{r} \right]_{r\rightarrow 0},
\end{equation}
which yields
\begin{equation}\label{u_st}
u_{\rm st}(\kappa,\Gamma) = \frac{\kappa(\kappa+1)\Gamma}{(\kappa+1)+(\kappa-1)e^{2\kappa}}-\frac{\kappa\Gamma}{2}-\frac{3\Gamma}{2\kappa^2}.
\end{equation}
The first (positive) term on the right-hand side of Eq.~(\ref{u_st}) corresponds to the particle-particle correlations in the ISM
approximation, the last two (negative) terms  represent the energy of the sheath around the particles and the energy of the neutralizing medium, respectively.

It is easy to demonstrate that in the limit $\kappa\rightarrow 0$, Eq.~(\ref{potential}) reduces to
\begin{displaymath}
\phi_{\rm OCP}(r)=\frac{Q}{r}+\frac{Q}{2a^3}r^2-\frac{9 Q}{5 a}.
\end{displaymath}
Combined this with Eq.~(\ref{energydef}) immediately yields the OCP result of Eq.~(\ref{uOCP}).

\begin{figure}
\includegraphics[width=7cm]{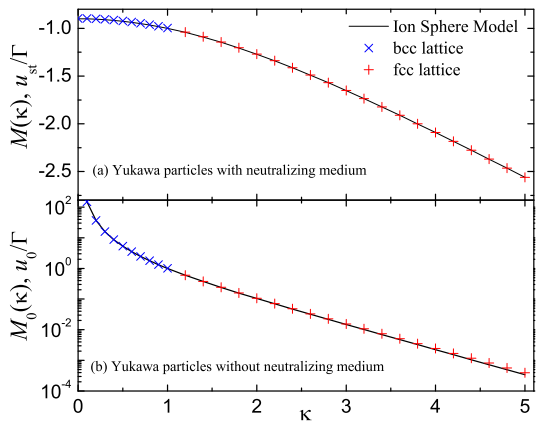}
\caption{(color online) Madelung constant of Yukawa systems as a function of the screening parameter $\kappa$. Two situations are shown: Particles with Yukawa interactions immersed in a neutralizing medium (a) and particles with Yukawa interactions without neutralizing medium, i.e. single component Yukawa systems (b). Symbols correspond to exact results for $M(\kappa)$ and $M_0(\kappa)$ taken from Refs.~\onlinecite{Farouki1994} and~\onlinecite{Hamaguchi}. Curves show $u_{\rm st}/\Gamma$ and  $u_0/\Gamma$ calculated using the ion sphere model.
}
\label{Madelung}
\end{figure}

Equation (\ref{u_st}) represents the {\it static} component of the excess energy of Yukawa systems within the framework of the ISM approximation, i.e. at strong coupling. Clearly, the energy is proportional to $\Gamma$. It is instructive to compare the ($\kappa$-dependent) coefficient of proportionality with the values of the Madelung constant, defined as
\begin{displaymath}
M(\kappa)=\lim_{\Gamma\rightarrow \infty}\frac{u(\kappa,\Gamma)}{\Gamma}.
\end{displaymath}
The exact values of the static energy and hence $M(\kappa)$ depend on the lattice type formed by the particles. Yukawa systems are known to form either bcc or fcc lattices in the equilibrium solid phase. In real dusty plasma experiments, the solid phase is often dominated by the hexagonal-close-packed (hcp) structures,~\cite{ISS1,ISS2,Mitic} which possibly indicates the non-equilibrium character of these systems or some deviations from the Yukawa interaction potential between the particles. Conventional Yukawa fluids freeze into the bcc solid in the regime of week screening and into the fcc solid at strong screening.~\cite{Hamaguchi,Robbins} The fcc-bcc-fluid triple point is located near $\kappa\simeq 4.5$.~\cite{Hamaguchi,Dupont,Hoy,KhrapakEPL2012} The difference between the values of Madelung constants for bcc and fcc solids is tiny in the regime of our interest, so we simply take the smallest value (which corresponds to bcc at $\kappa\lesssim 1$ and to fcc otherwise) to compare with the structure-independent value given by the ISM approximation [Eq.~(\ref{u_st})].
This comparison is shown in Fig.~\ref{Madelung}(a). The agreement is excellent. Relative deviations between the exact and ISM results amount to less than $\simeq 0.5\%$ at $\kappa=0$, to $\simeq 0.3\%$ at $\kappa = 1$ and diminish to $\simeq 0.003\%$ at $\kappa = 5$.

The excellent agreement between the exact results and the ISM approximation at large $\kappa$ should be considered with some care. In this regime, the dominant contribution to the Madelung energy is associated with the sheath-particle interactions ($M\simeq -\kappa/2$). Nevertheless, ISM approximation predicts rather accurately the internal energy coming from the particle-particle correlations, too. In order to illustrate this we subtract the energy associated with the neutralizing medium in Eq.~(\ref{u_st}). The remaining part,
\begin{align}\label{f0}
\nonumber u_0(\kappa,\Gamma)=f_0(\kappa)\Gamma, \\
f_0(\kappa)=  \frac{\kappa(\kappa+1)}{(\kappa+1)+(\kappa-1)e^{2\kappa}},
\end{align}
would correspond to the internal energy of the {\it single component} Yukawa system, i.e. to an imaginary system of particles interacting via the repulsive Yukawa potential without any neutralizing medium. The Madelung constant $M_0(\kappa)$ of this system should be compared with the ISM estimate $f_0(\kappa)$. Comparison shown in Fig.~\ref{Madelung}(b) demonstrates that the agreement is again excellent, but only in the weak screening regime, $\kappa\lesssim 1$. Deviations then increase, reaching $\simeq 1\%$ at $\kappa=2$ and $\simeq 10\%$ at $\kappa=4$.

To conclude this Section we summarize the difference between the DHH and ISM approximations. In the DHH, the cell (hole) radius $h$ is not fixed. It is determined from the boundary condition $\phi (h)=T/Q$ (which implies that the linearized particle density vanishes at the hole boundary) along with the requirement that the electrical potential and its derivative are continuous at the hole boundary. The cell is not charge neutral. In the ISM approximation, the cell radius is fixed and equal to the Wigner-Seitz radius, the cell is charge neutral. In the Appendix~\ref{APY} we show that the ISM result of Eq.~(\ref{f0}) can be also obtained from the Percus-Yevick (PY) radial distribution function for hard spheres. This provides the relation between the ISM approximation and the integral equation theories.

\section{Thermal corrections and total excess energy}

In Sec. II we have calculated the static energy of Yukawa systems within the framework of the ISM approximation. This assumes that the particle is located at the center of the ion sphere. However, thermal motion may result in some deviations of the particle position from the center and hence in some corrections to the internal energy. To estimate these we follow the procedure that has been recently suggested in the context of the (two- and three-dimensional) OCP.~\cite{23OCP} We calculate the change of the potential energy of the particle as a function of the distance from the cell center. Subtracting from Eq.~(\ref{potential}) the self-potential of the particle and irrelevant constant terms we get the variation of electrical potential with distance from the center $\delta \phi(r)$. The position-dependent energy associated with particle deviations is evaluated as $\delta W(r)= Q\delta \phi(r)$. This yields
\begin{equation}\label{Wr}
\delta W(r)/T= 2\Gamma f_0(\kappa)\left[\frac{\sinh(k_{\rm D}r)}{k_{\rm D}r}-1\right].
\end{equation}
To get thermal corrections we then average $\delta W(r)/T$ over the classical Gibbs distribution,
\begin{equation}\label{u_th}
u_{\rm th}(\kappa,\Gamma) = \frac{\int_0^a \delta W(r) r^2  e^{-\delta W(r)/T}dr}{T \int_0^a r^2 e^{-\delta W(r)/T} d r},
\end{equation}
where the integration is over the Wigner-Seitz cell volume. The dependence on $\kappa$ and $\Gamma$ comes from $\delta W(r)$.
In the OCP limit ($\kappa=0$), simple analytical expression for $u_{\rm th}$ can be derived,~\cite{23OCP} otherwise numerical integration is required.
The thermal contribution $u_{\rm th}$ calculated in this way tends to $3/2$ in the limit of very strong coupling ($\Gamma\rightarrow \infty$), as expected.~\cite{Ichimaru,Itoh}
As $\Gamma$ decreases, $u_{\rm th}$ demonstrates monotonous decrease.

By construction, the outlined approach is clearly more suitable for the solid phase and in fact it has much in common with that of Ref.~\onlinecite{Mansoori}, developed for the calculation of the thermodynamic properties of the solid phase and fluid-solid phase equilibria. Moreover, it is known that the actual excess thermal energy behaves non-monotonously and even exhibits a discontinuity (along with the excess entropy) at the fluid-solid phase transition, $\Delta u_{\rm th}\simeq 0.7$ for weakly screened Yukawa systems.~\cite{Rosenfeld95,Hamaguchi96} Nevertheless, we show below that this approach yields reasonable agreement with the exact data from numerical simulations, both in the solid phase and relatively deep into the fluid phase, although it does not reproduce the exact behavior of $u_{\rm th}$ in the vicinity of the fluid-solid phase transition.

\begin{figure}
\includegraphics[width=7.5cm]{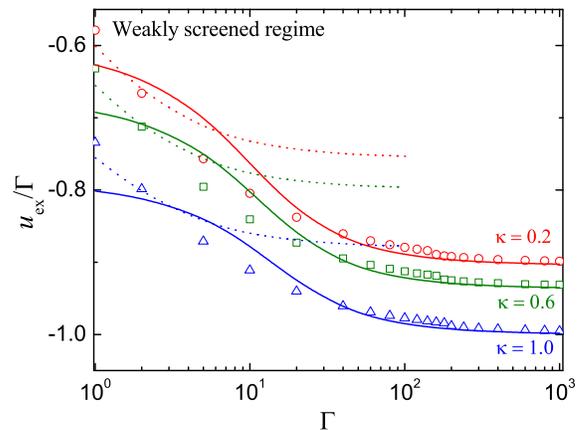}
\caption{(color online) Reduced excess energy (in units of $\Gamma$) as a function of the coupling parameter $\Gamma$ in the regime of weak screening, $\kappa\leq 1$. Solid curves correspond to the ISM approximation. Dotted curves show the result of the DHH approach.~\cite{DHH} Symbols are the exact results from MD simulations.~\cite{Farouki1994} Data for $\kappa=0.2$, $0.6$, and $1.0$ are shown. The OCP limit ($\kappa=0$) has been considered previously.~\cite{23OCP}}
\label{Uex1}
\end{figure}

\begin{figure}
\includegraphics[width=7.5cm]{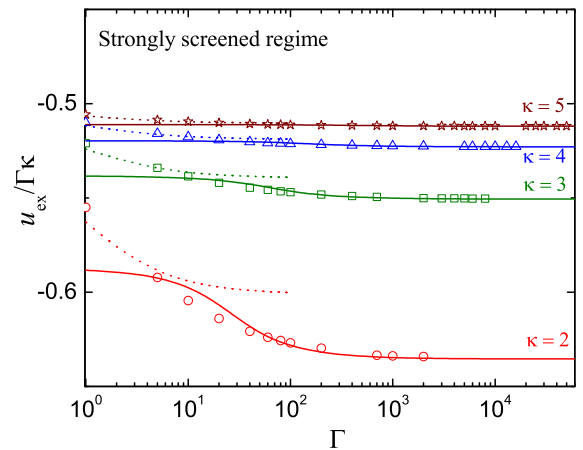}
\caption{(color online) Reduced excess energy (in units of $\Gamma\kappa$) as a function of the coupling parameter $\Gamma$ in the regime of strong screening, $\kappa > 1$. Solid curves correspond to the ISM approximation. Dotted curves show the result of the DHH approach.~\cite{DHH} Symbols are the results from MD simulations.~\cite{Hamaguchi} Data for $\kappa=2.0$, $3.0$, $4.0$ and $5.0$ are shown.}
\label{Uex2}
\end{figure}

The total excess energy is composed of the static and thermal contributions,
\begin{displaymath}
u_{\rm ex}= u_{\rm st}+u_{\rm th}.
\end{displaymath}
In our present approach, the static component is calculated from Eq.~(\ref{u_st}) and the thermal component from Eq.~(\ref{u_th}).
Comparison between these calculations and exact numerical results is shown in Fig.~\ref{Uex1} and~\ref{Uex2} for the regime of weak and strong screening, respectively.
The quantitative agreement is good at strong coupling, qualitative agreement is preserved down to $\Gamma\sim {\mathcal O}(1)$.

Dotted lines in Figs.~\ref{Uex1} and \ref{Uex2} show the results from our previous paper,~\cite{DHH} describing the application of DHH approach to Yukawa systems.
Comparison with ISM demonstrates the complementarity of these two simple approaches. The DHH approximation can describe accurately the regime from very weak (where it reduces to the conventional DH theory) to moderate coupling, $\Gamma\sim {\mathcal O}(1)$, while the ISM approximation is superior to DHH at strong coupling. At strong screening, the results from these approaches are hardly distinguishable. This does not imply that they coincide, but rather the excess energy is dominated by the sheath-related contribution.

\section{Relation to the Einstein frequency}

\begin{figure}
\includegraphics[width=7cm]{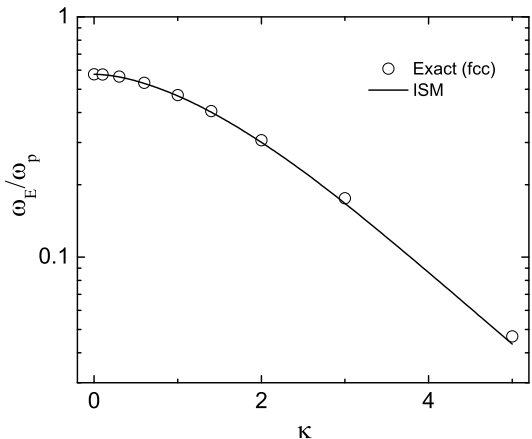}
\caption{Ratio of the Einstein to plasma-particle frequency, $\omega_{\rm E}/\omega_{\rm p}$, as a function of the screening parameter $\kappa$. Symbols correspond to the exact results for the fcc lattice.~\cite{OhtaPoP2000} The curve is computed using the ISM approximation, Eq.~(\ref{wE}).}
\label{Einstein}
\end{figure}

The Einstein frequency $\omega_{\rm E}$ is the characteristic oscillation frequency of a particle about its equilibrium position in a given crystalline lattice, when all other particles are located in their lattice sites. In the ISM approximation we can expand $\delta W (r)$, given by Eq.~(\ref{Wr}), into power series around $r=0$. The first term of this expansion is quadratic in $r$ and is proportional to the squared Einstein frequency, $\delta W(r)\simeq \tfrac{1}{2}m_{\rm p}\omega_{\rm E}^2r^2$, where $m_{\rm p}$ is the particle mass. For the Yukawa potential the calculation of $\omega_{\rm E}$ is particulary simple since it is trivially related to the Madelung constant of the single component Yukawa system.~\cite{Robbins} The result of the ISM approximation can be written as
\begin{equation}\label{wE}
\omega_{\rm E}^2=\frac{2}{9}\omega_{\rm p}^2\kappa^2f_0(\kappa),
\end{equation}
where $\omega_{\rm p}=\sqrt{4\pi Q^2 n_{{\rm p}0}/m_{\rm p}}$ is the plasma frequency associated with the particle component.
The ratio of $\omega_{\rm E}/\omega_{\rm p}$, as a function of $\kappa$, is shown in Fig.~\ref{Einstein}. Symbols correspond to the exact results, the curve is computed from Eq.~(\ref{wE}).
The latter is essentially exact in the OCP limit $\kappa\rightarrow 0$. Using $1+\kappa+(\kappa-1)e^{2\kappa}\simeq \tfrac{2}{3}\kappa^2+\mathcal{O}(\kappa^4)$ for small $\kappa$ we get
$\omega_{\rm E}/\omega_{\rm p}\simeq 3^{-1/2}$ as it should be for the OCP. As $\kappa$ increases, deviations between ISM and exact results grow. At $\kappa=3.0$ relative deviation is $\simeq 2\%$, while at $\kappa=5.0$ it amounts to $\simeq 7\%$. The ISM approximation underestimates the actual value of the Einstein frequency.

In the ISM approximation the Einstein frequency does not depend on the crystalline structure since the difference in Madelung constants between different lattices is not resolved. Note a trivial relation between $\omega_{\rm E}$ and the corresponding $M$ for particles immersed in the neutralizing medium
\begin{displaymath}
\omega_{\rm E}^2=\frac{2}{9}\omega_{\rm p}^2\kappa^2\left(M+\frac{\kappa}{2}+\frac{3}{2\kappa^2}\right).
\end{displaymath}
This can be used to evaluate $\omega_{\rm E}$ when the value of $M$ for a particular lattice is known.

The Einstein frequency can be approximately related to the well-known Lindemann criterion of melting.~\cite{Lindemann} This criterion states that a crystalline solid melts when the root-mean-square displacement of particles about their equilibrium lattice positions exceeds a certain fraction of the characteristic nearest neighbor distance. The critical fraction, known as the Lindemann parameter $L$, is expected to be a quasiuniversal quantity. In fact, however, its exact value may depend on such factors as crystalline structure and nature (shape) of the interparticle interactions.~\cite{Agrawal,Saija} For our present purposes we define the Lindemann-like parameter as the mean square deviation of the particle from the center of the cell, normalized to the cell radius, \begin{equation}\label{L}
L=\sqrt{\langle \delta r^2\rangle/a^2}.
\end{equation}
This is different from the conventional definition, because the Wigner-Seitz radius is used rather than nearest-neighbor spacing. In the Einstein approximation we have
\begin{equation}\label{L1}
\langle \delta r^2 \rangle \simeq \frac{3T}{m_{\rm p}\omega_{\rm E}^2}.
\end{equation}
Within the ISM, this should be adequate near the OCP limit because the potential energy of particle deviation from the cell center is exactly harmonic,~\cite{23OCP} $\delta W(r)\propto r^2$.
For stronger screening this can be less appropriate and we employ averaging over the Gibbs distribution
\begin{equation}\label{L2}
\langle \delta r^2 \rangle =  \frac{\int_0^a r^4 e^{-\delta W(r)/T}dr}{ \int_0^a r^2 e^{-\delta W(r)/T} d r}.
\end{equation}
In the OCP limit Eqs.~(\ref{L1}) and (\ref{L2}) are identical, but they can differ at non-zero $\kappa$.

\begin{figure}
\includegraphics[width=7cm]{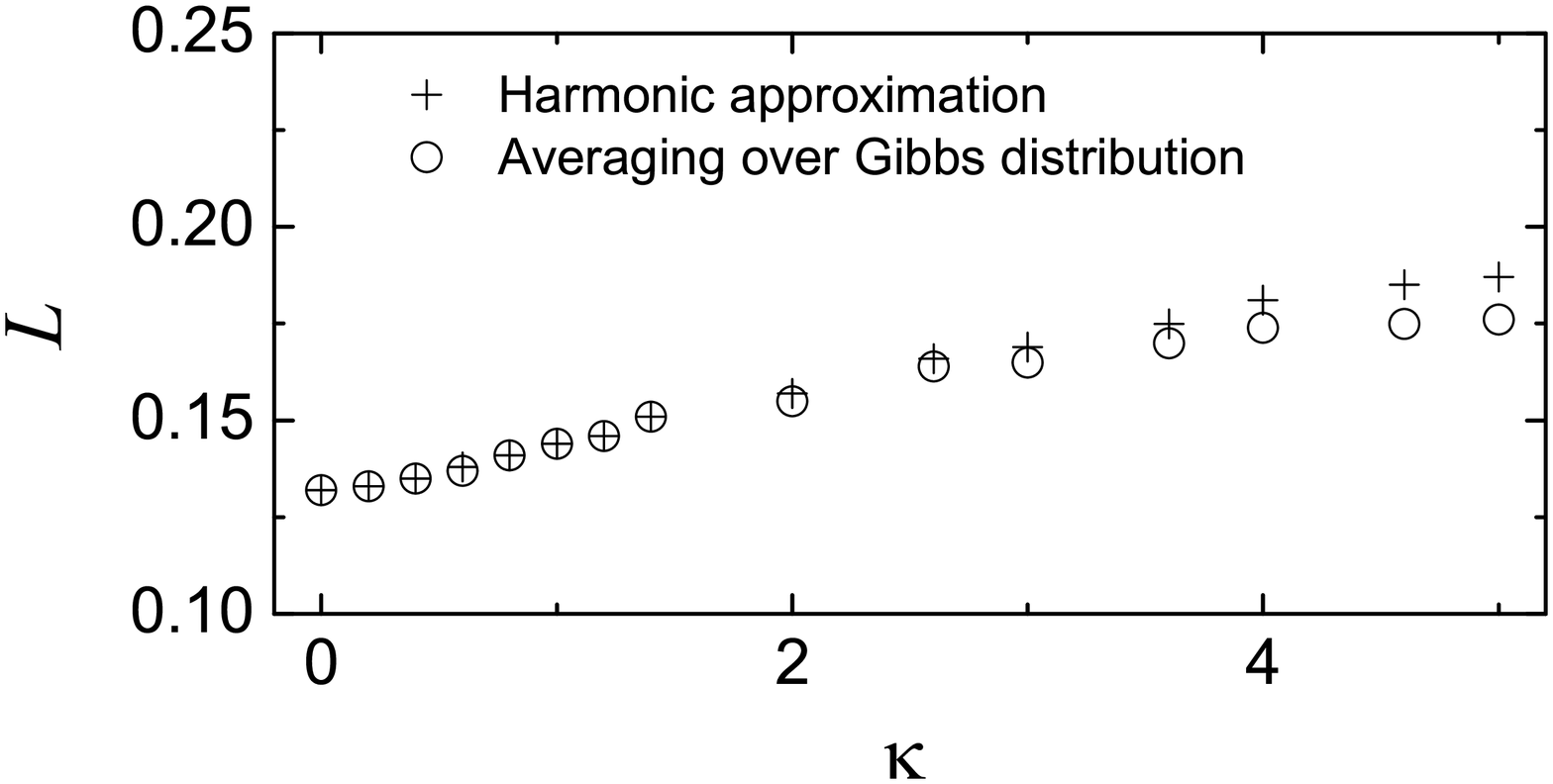}
\caption{Values of the Lindemann parameter $L$ defined in Eq.~(\ref{L}) at the fluid-solid phase transition of Yukawa systems vs. the screening parameter $\kappa$. Crosses correspond to the Einstein approximation [Eq.~(\ref{L1})], circles to the averaging over the Gibbs distribution [Eq.~(\ref{L2})]. Calculations make use of the values $\Gamma_{\rm melt}(\kappa)$ tabulated in Ref.~\onlinecite{Hamaguchi}.}
\label{Lindemann}
\end{figure}

We have plotted the values of $L$ at the fluid-solid phase transition evaluated from Eqs.~(\ref{L1}) and (\ref{L2}) in Figure~\ref{Lindemann}.
This figure shows that $L$ is not constant at melting of Yukawa systems, although it is rather weak function of $\kappa$. The Einstein approximation is practically indistinguishable from the full averaging procedure at $\kappa\lesssim 2$. For larger $\kappa$ deviations become observable, but remain relatively small. The qualitative behavior in the dependence of $L$ on the potentials steepness (i.e. $\kappa$) can be compared with that reported for the inverse-power-law potentials.~\cite{Agrawal} There, the evaluated (true) Lindemann parameter has a maximum near the power $\sim 6$ and somewhat decreases towards both soft and hard interaction limits (see Fig. 8 of Ref.~\onlinecite{Agrawal}). This is consistent with what we see in Fig.~\ref{Lindemann}, except the decrease of $L$ at large $\kappa$ is not confirmed due to the lack of the accurate data for $\kappa >5 $. Note that a related parameter -- the localization length at the glass transition of Yukawa systems -- has been recently shown to reach a maximum around $\kappa\simeq 10$.~\cite{Yazdi} Similar tendency can be expected for $L$ since the glass-transition and melting lines for the Yukawa potential are essentially parallel in the $(\kappa, \Gamma)$ plane in a rather wide range of $\kappa$.~\cite{Yazdi,Sciortino}

From the quantitative point of view, however, neither Eq.~(\ref{L1}) nor (\ref{L2}) is very useful to predict the actual value of $\langle \delta r^2 \rangle$. It is well known that to get the correct result in the quasi-harmonic approximation, the Einstein estimate (\ref{L1}) should be multiplied by a factor $\omega_{\rm E}^2\langle\omega^{-2}\rangle$, where the averaging is over all phonon wave vectors and polarizations. For the Yukawa interactions this ($\kappa$-dependent) factor lies in the range between $\simeq 4$ and $\simeq 2$ (for $1\lesssim \kappa\lesssim 10$) as documented in Ref.~\onlinecite{Robbins} and the actual values of $L$ at melting are expected to be around $0.26$ (bcc solid) and $0.27$ (fcc solid).~\cite{Dubin1994} Since Eq.~(\ref{L2}) does not demonstrate any improvement compared to the simple Einstein approximation, but do involves numerical integration, it is not very useful for practical applications.

\section{Pressure and compressibility}

We return to the thermodynamic properties of Yukawa systems and derive an equation for the excess pressure and compressibility of the particle component in the ISM approximation. As has
been discussed,~\cite{DHH} the excess (free) energy of the sheath does not contribute to the excess pressure. The
excess pressure arising from the particle-particle correlations can be conveniently evaluated from the virial pressure
equation involving the radial distribution function $g(r)$ of the particle component.~\cite{HansenBook} Taking into account the neutralizing medium we can write for the excess pressure~\cite{DHH}
\begin{equation}
p_{\rm ex}=-\frac{2\pi n_{{\rm p}}}{3T}\int_0^{\infty}r^3V^{\prime}(r)\left[g(r)-1\right]dr,
\end{equation}
where $V(r)=(Q^2/r)\exp(-k_{\rm D}r)$ is the Yukawa pair interaction potential.
We have also an expression relating the excess energy and $g(r)$,
\begin{equation}\label{energy}
u_{\rm ex}=\frac{2\pi n_{{\rm p}}}{T}\int_0^{\infty}r^2V(r)\left[g(r)-1\right]dr - \frac{\kappa\Gamma}{2}.
\end{equation}
From these two equations we obtain a very useful approximate relation between the excess pressure and energy of Yukawa systems,
\begin{equation}\label{Rel}
p_{\rm ex}=\frac{1}{3}\left(u_{\rm ex}-\kappa\frac{\partial u_{\rm ex}}{\partial \kappa}\right).
\end{equation}
This approximation is valid for both single component and conventional Yukawa systems with neutralizing medium [note that the term $-\kappa\Gamma/2$ in $u_{\rm ex}$ automatically cancels out when substituted in Eq.~(\ref{Rel})]. In deriving (\ref{Rel}) we neglected the dependence of $g(r)$ on $\kappa$. This assumption is accurate in the weakly screened regime (see e.g. Fig. 6 of Ref.~\onlinecite{Farouki1994}), but possibly less justified at stronger screening. In the strongly coupled regime, however, the excess energy is dominated by the static contribution, which is given by Eq.~(\ref{u_st}) in the ISM approximation. If only static contribution is retained, then Eq.~(\ref{Rel}) is thermodynamically consistent and results in a simple analytical expression for the excess pressure
\begin{equation}\label{pex}
p_{\rm ex}(\kappa,\Gamma)\simeq \frac{\kappa^4\Gamma}{6\left[\kappa\cosh(\kappa)-\sinh(\kappa)\right]^2}-\frac{3\Gamma}{2\kappa^2}.
\end{equation}
The first (positive) term describes particle-particle correlations. It corresponds to the excess pressure of an imaginary single component Yukawa system.
The second (negative) term represents the contribution of the neutralizing medium. This contribution is responsible for the negative sign of the excess pressure of strongly coupled Yukawa systems.

Near the OCP limit ($\kappa\rightarrow 0$) we have the following expansion of $p_{\rm ex}$ in powers of $\kappa$:
\begin{multline}\label{pOCP}
p_{\rm ex}(\kappa,\Gamma)\simeq -\tfrac{3}{10}\Gamma + \tfrac{6}{175}\kappa^2\Gamma -\tfrac{47}{15750}\kappa^4\Gamma\simeq \\ -\Gamma\left(0.3-0.034\kappa^2+0.003\kappa^4\right).
\end{multline}
Equation (\ref{pOCP}) indicates that the pressure {\it increases} with $\kappa$ in agreement with Refs.~\onlinecite{Faussurier} and~\onlinecite{DHH} and disagreement with Ref.~\onlinecite{Farouki1994}. This difference is the result of different assumptions regarding the relation between the density of the particle component and neutralizing medium.

As a check of the accuracy of Eq.~(\ref{pOCP}) we consider the leading term in the excess energy dependence on $\Gamma$, $u_{\rm ex}\simeq a(\kappa)\Gamma$.
A very accurate fit for $a (\kappa)$, based on the MD numerical results, has been suggested,~\cite{Hamaguchi96} $a(\kappa)\simeq -0.899-0.103\kappa^2+0.003\kappa^4$. Substituting this into Eq.~(\ref{Rel}) we get after simple algebra
$p_{\rm ex}\simeq -\Gamma(0.300-0.034\kappa^2+0.003\kappa^4)$, in excellent agreement with Eq.~(\ref{pOCP}).

The quantity which is often of interest when dealing with hydrodynamic description of wave phenomena in strongly coupled dusty plasmas~\cite{DHH,Kaw1998,Kaw2001,Salin} is the inverse reduced isothermal compressibility, $\mu_{\rm p}=(1/T)(\partial P/\partial n_{\rm p})_{T}$. It is related to the excess pressure via
\begin{equation}\label{compr}
\mu_{\rm p}= 1+p_{\rm ex}+\frac{\Gamma}{3}\frac{\partial p_{\rm ex}}{\partial \Gamma}-\frac{\kappa}{3}\frac{\partial p_{\rm ex}}{\partial \kappa}.
\end{equation}
Substituting Eq.~(\ref{pex}) for $p_{\rm ex}$ we get a simple and practical expression for $\mu_{\rm p}$,
\begin{equation}\label{mu1}
\mu_{\rm p}(\kappa,\Gamma)\simeq 1-\frac{3\Gamma}{\kappa^2}+\frac{\Gamma\kappa^6\sinh(\kappa)}{9\left[\kappa\cosh(\kappa)-\sinh(\kappa)\right]^3}.
\end{equation}
Near the OCP limit series expansion of Eq.~(\ref{mu1}) yields
\begin{displaymath}
\mu_{\rm p}(\kappa,\Gamma)\simeq 1-\tfrac{2}{5}\Gamma+\tfrac{4}{175}\kappa^2\Gamma+{\mathcal O}(\kappa^6\Gamma).
\end{displaymath}

\begin{table}
\caption{\label{Tab0} Inverse reduced isothermal compressibility, $\mu_{\rm p}$, of Yukawa systems for several strongly coupled state points, for which dispersion relations have been obtained in a numerical experiment.~\cite{DAW1,DAW2} Results of calculations using various approximations (OCP, DHH, ISM, NN, SMSA) are summarized. For details see the text.}
\begin{ruledtabular}
\begin{tabular}{lllllll}
$\kappa$ & $\Gamma$ & OCP &  DHH & ISM & NN & SMSA  \\ \hline
0.3 & 144 & -55.5 & -46.3 & -56.3 &  --   & -54.9\\
1.0 & 207 & -80.5 & -59.4 & -77.1 & -294.3   & -75.2\\
2.0 & 395 & -156.8 & -84.8 & -123.2 & -134.7  & -119.4 \\
3.0 & 1100 & -- & -168.9 & -257.1 & -256.2 & -249.5 \\
\end{tabular}
\end{ruledtabular}
\end{table}

In Ref.~\onlinecite{DHH} we have described the simplistic hydrodynamic model of the dust acoustic waves (DAW) in strongly coupled dusty plasmas (modeled by Yukawa systems). This model yields the dispersion relation of the form
\begin{equation}\label{DR}
\frac{\omega^2}{\omega_{\rm p}^2}=\frac{q^2}{q^2+\kappa^2}+\frac{q^2}{3\Gamma}\mu_{\rm p},
\end{equation}
where $\omega$ is the wave frequency, $q=ka$ is the reduced wavenumber, and the adiabatic index is set unity at strong coupling. The strong coupling effects come into (\ref{DR}) only via $\mu_{\rm p}$. When compared with the dispersion relations obtained from MD simulations~\cite{DAW1,DAW2} for several strongly coupled state points near freezing, the theory demonstrated reasonable accuracy.~\cite{DHH} The values of $\mu_{\rm p}$ have been calculated using the DHH approximation and it makes sense now to compare these with the more accurate results from the ISM approximation and other relevant approaches.

The comparison between different approaches is shown in Table~\ref{Tab0}. Here the first two columns contain the values of $\kappa$ and $\Gamma$, characterizing the state of the system (all the state points are rather close to the fluid-solid phase transition, see Fig. 4 from Ref.~\onlinecite{DHH}). The third column shows the compressibility of the OCP at a given $\Gamma$, as calculated using the fitting formula for the OCP excess energy.~\cite{Hamaguchi96} It provides reasonable estimate of the compressibility of Yukawa systems at $\kappa\lesssim 1$ and should clearly not be used at $\kappa\gtrsim 2$. The fourth column shows the values obtained using the DHH approximation.~\cite{DHH} The next column gives the values obtained within the framework of the ISM approximation [Eq.~(\ref{mu1})]. One more column shows the values obtained using the nearest neighbor (NN) approximation (discussed in the Sec. VI) using the three-term version of Eq.~(\ref{pNN}). This should not be applied for $\kappa\lesssim 2$, but becomes progressively more and more accurate as $\kappa$ increases. The last column provides the value obtained with the help of the soft mean spherical approximation (SMSA) proposed very recently.~\cite{SMSA} This integral theory approach is thermodynamically consistent and demonstrates remarkable agreement with the thermodynamic quantities from MD simulations. We expect, therefore, that the values from SMSA lie most closely to the actual ones (among listed in Table~\ref{Tab0}). This implies that the DHH approach overestimates the actual compressibility by some $\simeq 15\%$ in the OCP limit and by $\simeq 30\%$ at $\kappa=3$. The values of $\mu_{\rm p}$ obtained using the ISM are only by few percent smaller than those from SMSA for all values of $\kappa$ considered. This is clearly a very good performance taking into account the simplicity of the ISM approach. Note that when calculating $\mu_p$  using the ISM and NN approximation we have neglected the thermal contribution to $u_{\rm ex}$. This demonstrates the relative magnitude of thermal contribution, which does not plays very significant role in the considered regime.

\begin{figure}
\includegraphics[width=8cm]{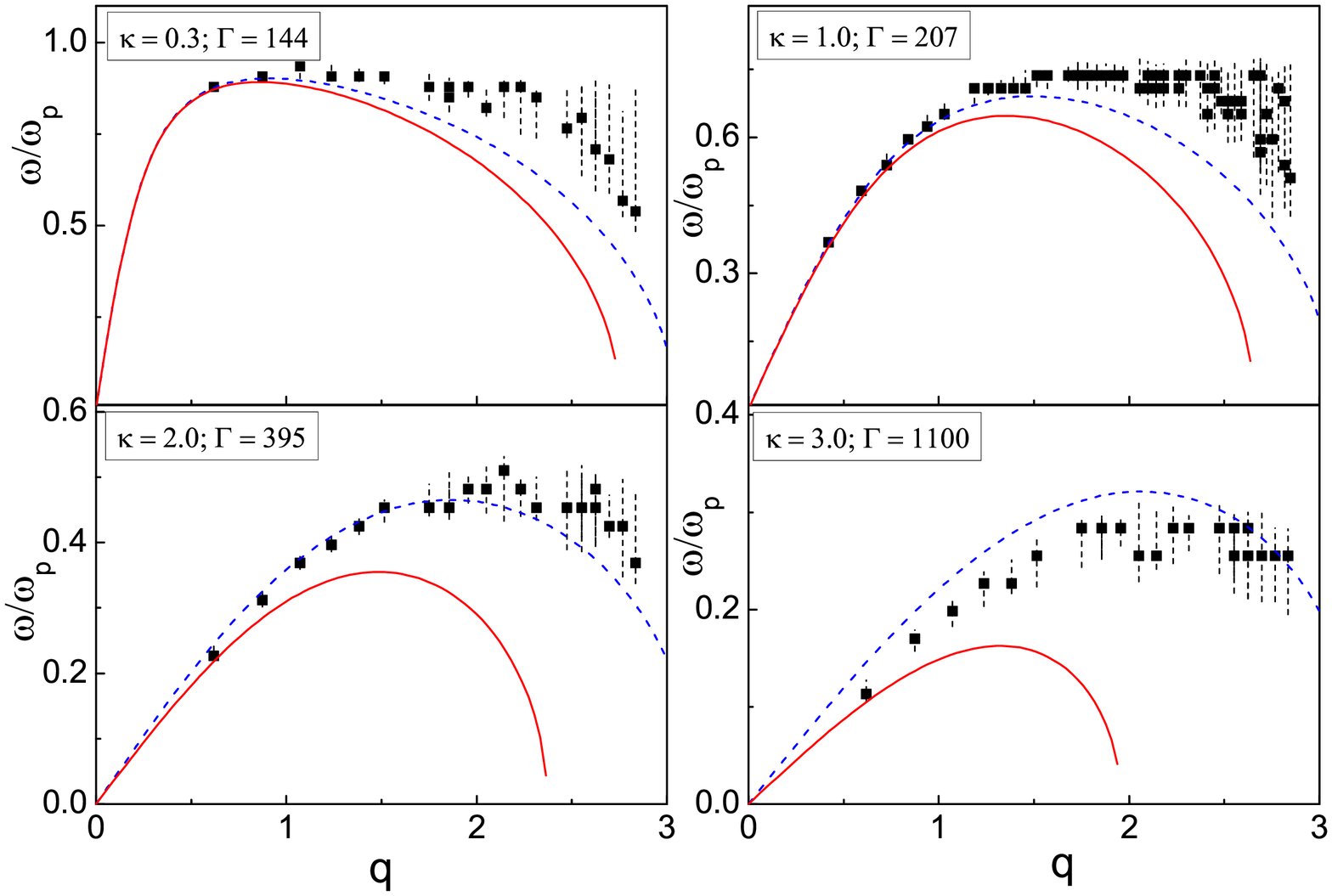}
\caption{(color online) Dispersion of the longitudinal waves in Yukawa fluids near freezing. Symbols correspond to the results from numerical experiment.~\cite{DAW1,DAW2} Solid (red) curves are calculated using Eq.(\ref{DR}) with the ISM values for $\mu_p$. Dashed (blue) curves are calculated using Eq.(\ref{DR}) with the DHH values for $\mu_p$.~\cite{DHH} Note that with the new more accurate values of $\mu_p$ the agreement with the numerical data becomes worse. For the discussion see the text.}
\label{DAW}
\end{figure}

The new dispersion curves, obtained by substituting ISM values for $\mu_{\rm p}$ into Eq.~(\ref{DR}) are compared with the results from numerical experiments in Figure~\ref{DAW}. In contrast to our expectations, more accurate values of the compressibility result in {\it worse} agreement with the numerical data. The deviations are particularly pronounced in the short-wavelength limit and they increase with $\kappa$. This finding is not completely surprising: The hydrodynamic description becomes progressively less justified when the wavelength becomes comparable to the interparticle distance. Among existing alternatives to describe wave-related phenomena in Yukawa systems  we can mention the quasi-localized charge approximation,~\cite{Rosenberg,Kalman,Donko,Rosenberg2014} which has been demonstrated to agree reasonably well with the numerically obtained dispersion relations at strong coupling.~\cite{DAW1,Kalman}

\section{Strong screening regime}

In the previous sections we have seen that the ISM approximation is remarkably accurate at weak screening (especially near the OCP limit), but becomes progressively less accurate as $\kappa$ increases. The fast exponential decay of the Yukawa potential in this regime gives hope that the nearest neighbor approximation can become a simple and reliable alternative to evaluate the internal   energy and other thermodynamic properties. Assuming perfect crystalline order and summing over the shells of neighbors around a test central particle, the reduced internal energy can be written as
\begin{equation}\label{NN}
u_0(\kappa,\Gamma)=\frac{1}{2}\Gamma\sum_i (N_i/z_i)\exp(-\kappa z_i),
\end{equation}
where $N_i$ is the number of neighbors within the $i$th shell, and $z_i$ is the radius of the $i$th shell (in units of $a$). Following Ref.~\onlinecite{Mansoori} we limit ourselves to three first shells.
The corresponding values of $N_i$ and $z_i$ for the fcc lattice (energetically favorable in the regime of strong screening) are summarized in Table~\ref{Tab}.

\begin{table}
\caption{\label{Tab} Values of $N_i$ and $z_i$ for the fcc lattice ($i\leq 3$).}
\begin{ruledtabular}
\begin{tabular}{lll}
$i$ & $N_i$ & $z_i$   \\ \hline
1 & 12 & 1.8094   \\
2 &  6 & 2.5589   \\
3 & 24 & 3.1340   \\
\end{tabular}
\end{ruledtabular}
\end{table}

\begin{figure}
\includegraphics[width=7cm]{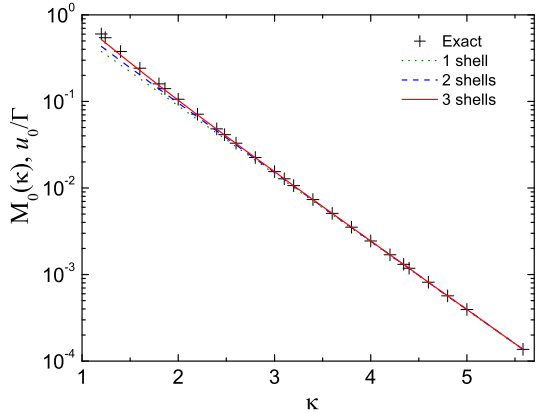}
\caption{(color online) Madelung constant of the single component Yukawa system forming the fcc lattice. Symbols correspond to exact results for $M_0(\kappa)$ tabulated previously.~\cite{Hamaguchi,Robbins} Curves are calculated using nearest-neighbor approximation [Eq.~(\ref{NN})] considering one (green dotted), two (blue dashed) and three (red solid) shells of the neighbors.}
\label{Madelung1}
\end{figure}

In Figure~\ref{Madelung1} the comparison between the actual Madelung constants, $M_0(\kappa)$, of the single component Yukawa systems forming the fcc solid and the values of $u_0/\Gamma$ calculated from Eq.~(\ref{NN}) is shown. Three curves are plotted, corresponding to including one, two, and three near neighbor shell into consideration.
The three-terms approximation provides the accuracy better than $1\%$ already at $\kappa\gtrsim 2.5$. Similar accuracy is reached for two- and one-term approximation at $\kappa\gtrsim 3.8$ and $\kappa\gtrsim 5.0$, respectively. Note that at $\kappa\gtrsim 5.0$ the relative difference between the three-term nearest neighbors approximation and the exact results is smaller than $10^{-5}$.
Comparing the nearest-neighbor approximation with that of the ISM we find that the three-term version of (\ref{NN}) provides better accuracy for $\kappa\gtrsim 2.1$.

Thermodynamic properties can be estimated from Eq.~(\ref{NN}) in cases when the static contribution to the excess energy dominates over the thermal one (strong coupling is a necessary condition for that). For instance, substituting Eq.~(\ref{NN}) into Eq.~(\ref{Rel}) immediately yields
\begin{equation}\label{pNN}
p_{\rm ex}\simeq \frac{\Gamma}{6}\sum_i(N_i/z_i)(1+\kappa z_i)\exp(-\kappa z_i)-\frac{3\Gamma}{2\kappa^2},
\end{equation}
where the last term is the contribution from the neutralizing medium, compare to Eq.~(\ref{pex}). The first term corresponds to the excess pressure of the single component Yukawa system. Note that since the structural properties are fixed in this near-neighbor approximation (the model radial distribution function has no dependence on $\kappa$), Eq.~(\ref{Rel}) becomes exact in this case.
When only $i=1$ term is retained, the result is equivalent, except the factor of two, to that used previously to estimate the pressure of crystalline dusty plasmas.~\cite{Gozadinos,Yaroshenko2010}
Note that the contribution from particle-particle correlations to the excess pressure decreases exponentially as $\kappa$ increases. At $\kappa=3$ its absolute value is $20\%$ of that due to neutralizing medium, while at $\kappa = 5.5$ it drops to $\simeq 1\%$.

\section{Conclusion}

We have discussed in detail the application of the ISM approximation to the idealized Yukawa systems, which can be of interest in the context of conventional plasmas, dusty (complex) plasmas, and colloidal dispersions. ISM provides a simple and efficient method to estimate the internal energy of these systems at strong coupling, both in the fluid and solid phases. The accuracy is not sufficient to make any predictions about the location of the fluid-solid phase transition, but is acceptable for many other purposes. For instance, simple analytical expressions for the excess pressure and inverse isothermal compressibility have been obtained in this paper.

ISM is reliable in the regime of strong coupling and weak screening. In this respect, is is complimentary to the DHH approximations that we discussed previously, which is applicable in the transitional regime between weak and moderate coupling. Moreover, as we have pointed out in the present paper, the nearest neighbor approximation provides quite good accuracy when the screening strength increases. Thus, various simple analytical approaches to estimate thermodynamic properties of Yukawa systems are available, in essentially entire range of possible phase states. In some cases, when idealizations behind the conventional Yukawa model are appropriate, these approaches can be applied to real systems (e.g. dusty plasmas and colloidal dispersions). Alternatively, they can serve as the basis of more realistic models, allowing an easy evaluation of the relative importance of specific system properties in various situations.

\begin{acknowledgments}
We would like to thank Panagiotis Tolias for providing us with the values of compressibility in the SMSA approximation~\cite{SMSA} listed in Table~\ref{Tab0} and Satoshi Hamaguchi for sending us the numerical results on longitudinal waves dispersion in Yukawa fluids shown in Fig. 6. The authors gratefully acknowledge the support of the Russian Science Foundation: Studies of thermodynamic properties of Yukawa systems have been supported by the Project No. 14-12-01235; studies of wave dispersion relations have been supported by the Project No. 14-43-00053.
\end{acknowledgments}

\appendix

\section{Relation to conventional three component dusty plasma}\label{ApA}

Normally, dusty plasma consists of three charged components: electrons, ions, and particles. Overall system quasi-neutrality implies $Qn_{{\rm p}0}+qn_{i0}-en_{e0}=0$, where $q$ is the ion charge (for singly charged ions $q=+e$). The electron and ion densities inside the cell satisfy $n_e=n_{e0}(1+e\phi/T_e)$ and  $n_i=n_{i0}(1-q\phi/T_i)$, where $T_e$ and $T_i$ are the corresponding temperatures, which are not necessarily equal. The Poisson equation inside the cell is
\begin{displaymath}
\Delta \phi = - 4\pi Q\delta({\bf r}) + k_{\rm D}^2\phi+4\pi Q n_{{\rm p}0},
\end{displaymath}
where $k_{\rm D}^2=4\pi (q^2n_{i0}/T_i+e^2n_{e0}/T_e)$ characterizes the inverse screening length. The general solution is given by Eq. (\ref{sol1}).  The conditions of cell neutrality and Coulomb asymptote at the origin then immediately results in potential distribution (\ref{potential}). Note that the model generally results in non-zero potential at the cell boundary (Wigner-Seitz radius).

\section{Equivalence of ISM and PY approximations}\label{APY}

The starting point is the energy equation for the single component Yukawa system,~\cite{HansenBook}
\begin{equation}\label{energy2}
u=\frac{2\pi n_{\rm p}}{T}\int_0^{\infty}r^2V(r)g(r)dr,
\end{equation}
where $g(r)$ is the radial distribution function and $V(r)$ is the Yukawa pair interaction energy. Using the PY radial distribution function $g_{\rm PY} (r)$ for hard spheres of diameter $d$ and packing fraction $\eta=\tfrac{\pi}{6}n_{\rm p}d^3$ we can rewrite Eq.~(\ref{energy2}) as
\begin{equation}
u_{\rm PY}=6\eta^{2/3}\Gamma\int_1^{\infty}xg_{\rm PY}(x)e^{-tx}dx=6\eta^{2/3}\Gamma G(t),
\end{equation}
where $x=r/d$, $t=k_{\rm D}d=2\eta^{1/3}\kappa$, and $g_{\rm PY} (x)$ also depends on $\eta$. The function $G(t)$ (for a given $\eta$) is known analytically~\cite{Wertheim,Thiele}
\begin{equation}
G(t)=\frac{tL(t)}{12\eta[L(t)+S(t)e^t]},
\end{equation}
where $L(t)= 12\eta[(1+\tfrac{1}{2}\eta)t+(1+2\eta)]$ and $S(t)=(1-\eta)^2t^3+6\eta(1-\eta)t^2+18\eta^2 t-12\eta(1+2\eta)$. The ion sphere model corresponds to $g(r)=0$ for $r\leq 2a$, that is we have to chose $d\equiv 2a$ or $\eta\equiv 1$ for consistency. This results in $L(2\kappa)=36(1+\kappa)$, $S(2\kappa)=36(\kappa-1)$, which immediately yields
\begin{equation}
u_{\rm PY}=\frac{\kappa(\kappa+1)\Gamma}{(\kappa+1)+(\kappa-1)e^{2\kappa}}=f_0(\kappa)\Gamma,
\end{equation}
in agreement with the result obtained from purely electrostatic consideration [Eq.~(\ref{f0})].

Note that similar (PY) estimation of the Madelung constants can be done for inverse-power-law potentials.~\cite{Rosenfeld1981}
Similarly to the case of Yukawa interactions, the accuracy is better for softer potentials.

\end{document}